\documentclass[a4paper,12pt]{article}
\pdfoutput=1 % if your are submitting a pdflatex (i.e. if you have
\usepackage{graphicx}
\usepackage{color}
\usepackage[colorlinks,hyperfootnotes=false]{hyperref}
\usepackage{fancyhdr}
\usepackage{amsfonts}
\usepackage{amssymb}
\usepackage{amsthm}
\usepackage{graphics}
\usepackage{multirow}
\usepackage{amsmath}
\usepackage{amsmath}
\usepackage{epsfig}
\usepackage{dcolumn}% Align table columns on decimal point
\usepackage{bm}% bold math

\usepackage{microtype}

\DeclareMathOperator{\arcsinh}{arcsinh}
\newcommand{\beq}{\begin{equation}}
\newcommand{\eeq}{\end{equation}}
\newcommand{\bea}{\begin{eqnarray}}
\newcommand{\eea}{\end{eqnarray}}

\newcommand{\lsim}{\mathrel{\hbox{\rlap{\lower.55ex\hbox{$\sim$}} \kern-.3em \raise.4ex \hbox{$<$}}}}
\newcommand{\gsim}{\mathrel{\hbox{\rlap{\lower.55ex\hbox{$\sim$}} \kern-.3em \raise.4ex \hbox{$>$}}}}
\newcommand{\drm}{\mathrm{d}}

\newcommand{\gam}{\Gamma_{\mathrm{mm}}}

\newcommand{\gamt}{\tilde{\Gamma}_{\mathrm{mm}}}
\newcommand{\rhos}{\rho_{\mathrm{mm}}}

\newcommand{\rhost}{\tilde{\rho}_\mathrm{mm}}
\newcommand{\rhort}{\tilde{\rho}_r}
\newcommand{\rhomt}{\tilde{\rho}_\mathrm{dm}}
\newcommand{\dels}{\delta_\mathrm{mm}}
\newcommand{\delr}{\delta_r}
\newcommand{\delm}{\delta_\mathrm{dm}}

\newcommand{\ther}{\theta_r}

\newcommand{\thest}{\tilde{\theta}_\mathrm{mm}}
\newcommand{\thert}{\tilde{\theta}_r}
\newcommand{\themt}{\tilde{\theta}_\mathrm{dm}}

\newcommand{\tilk}{\tilde{k}}

\begin{document}

\title{Structure Formation during an early period of matter domination}

\author{
Gabriela Barenboim and Javier Rasero\\[2ex]
\textit{Departament de F\'{\i}sica Te\`orica and IFIC},\\ \textit{Universitat de 
Val\`encia-CSIC, E-46100, Burjassot, Spain.}\\
	%E-mail: \email{gabriela.barenboim@uv.es}}
}

\maketitle
%\preprint{}	% OR: \preprint{Aaaa/Mm/Yy\\Aaa-aa/Nnnnnn}
			  	% Use \hepth etc. also in bibliography.  
\begin{abstract}
 In this work we show that modifying the thermal history of the Universe by including an early period of matter domination can lead to the formation of astronomical objects. However, the survival of these objects can only be possible if the dominating matter decays to a daughter particle which is not only almost degenerate with the parent particle but also has an open annihilation channel. This requirement translates in an upper bound for the coupling of such a channel and makes the early structure formation  viable.
\end{abstract}

%\keywords{Structure formation, Dark Matter, Physics of the Very Early Universe}
%\thispagestyle{fancy}
\newpage
\section{Introduction}
The Universe known and observed nowadays is a consequence of a long process where the primordial seeds were amplified due to Inflation, a stage of the Universe where its size grew exponentially and left all the observable scales out of the horizon. At re-entry after Inflation termination, the seeds of these scales began to accrete matter and formed the observable astrophysical structures. One of the advantages of the standard cosmological scenario is that it is capable of addressing the whole process since the beginning until the late formation of complex structures.

The first seeds are widely supposed to be quantum fluctuations, amplified during the primordial inflationary era. Working in the fourier space, the component responsible for conducting the early exponential expansion of the Universe develops an inhomogeneous perturbation with a certain length and amplitude that gets frozen when the horizon scale becomes smaller than this length. Such a perturbation is then transmitted to the other components of the Universy by gravity. Furthermore, the amplitude of this perturbation is nearly the same for every component of the Universe, once one assumes that perturbations are adiabatic, as experiments seem to confirm.

Inflation is commonly assumed to be followed by a radiation era once pressure becomes important against gravity. Due to this effect, seeds are unable to attract matter and therefore form structures. As a consequence, the gravitational potential for scales entering the horizon throughout this radiation dominated epoch vanishes, a feature exhibited by the power spectrum for such scales getting suppressed by a factor $\frac{1}{k^3}$.

Since the energy density of radiation is diluted with the expansion of the Universe more rapidly than the energy density of matter, pressure becomes insignificant after the matter-radiation equality point and structures can be formed by matter accretion. At this matter dominated stage, perturbations that enter the horizon start to grow linearly with the scale, attracting more matter until they become non linear and collapse into the observed structures. This effect can be seen in the power spectrum profile of the perturbations, that grows as $k$. One peculiarity of this whole process is that as a consequence of the coupling between baryons and photons, observable matter starts to fall into the gravitational wells at $z_{\rm dec}\approx 1100$, much later than dark matter which, as being weakly interacting, starts to grow and form structures right after the matter-radiation equality time $z_{\rm eq} \approx 3400$. This is why first and older objects are searched in the form of halos or mini-halos of dark matter. 

\begin{figure}[t]
\centering
 \includegraphics[scale=0.9]{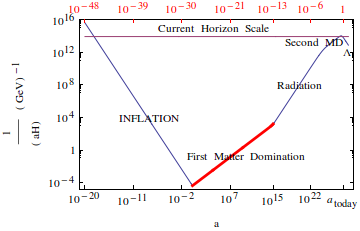}
 % thermal history.png: 794x1123 pixel, 96dpi, 21.00x29.70 cm, bb=0 0 595 842
 \caption{Evolution of the Hubble horizon in a non standard history of the Universe as a function of the scale factor. Scales factors in black (bottom of the plot) correspond to the convention used in this work where $a=1$ signals the beginning of matter domination. Scale factors in red (top of the plot) correspond to the convention where $a=1$ is set to today. This double labelling can be used as a ``dictionary'' for the following plots.}
 \label{fig:scaleev}
\end{figure}

Needless to say, this is the cartoon picture of structure formation assuming the standard thermal history of the Universe. However, as far as the Universe is radiation dominated by BBN, $T_{BBN}\simeq 1 \ {\rm MeV}$, and matter-radiation equality takes places at $T_{\rm eq} \simeq 1 \ {\rm eV}$, one is free to modify the thermal history at will. For instance, thermal inflation \cite{Lyth:1995ka} introduces a very short inflationary epoch to get rid of unwanted particles such as moduli.  Another case would be to consider that a very heavy particle, with a large thermal abundance, came to dominate the energy density of the Universe at its early stages. This early domination can be healthy for erasing unwanted particles and relaxing the conditions for producing the baryon asymmetry at the electroweak scale \cite{Barenboim:2012nh,Barenboim:2008zk}. An example of this latter modification in the standard history of the Universe can be seen in figure \ref{fig:scaleev}, that shows the evolution of the comoving Hubble radius through different eras. Within this sort of picture, an early matter domination era (coloured in red) driven by a heavy particle commences shortly after inflation ends, contrary to the standard picture where inflation and dark matter domination epochs are connected by a long period of radiation domination. One of the most remarkable changes when including such a modification is that scales entering the horizon during this new era can now grow linearly with the scale until their amplitudes become non linear and begin to form substructures, which in principle are to survive up to now. In this work, we study the conditions under which such a scenario can be realised and explore the consequences of such an early structure formation period in a thermal history of the Universe such as the one depicted in figure \ref{fig:scaleev}.

Our paper will be organised as follows: In section 2, we introduce the setup of the Universe, {\it i.e}, its components, interactions and magnitudes. Once the setup for such an Universe is given, the history that follows is automatically known. Furthermore, the details of the construction and motivation for such a scenario are also introduced. In section 3, the features of the structure formation picture are explained. We finally conclude in section 4.

\section{Scenario details}
In \cite{Erickcek:2011us}, a multifluid Universe where a heavy matter particle dominates the thermal history of the Universe until it decays away into radiation and matter was considered. The equation of motions for this case are 
\bea
\frac{\drm\rho_{\rm mm}}{\drm t}  +3H\rho_{{\rm mm}} &=& -\gam \rhos \;,\\
\frac{\drm\rho_r}{\drm t} +4H\rho_r &=& (1-f_b)\gam \rhos \;, \\
\frac{\drm\rho_{dm}}{\drm t} +3H\rho_{dm} &=& f_b\gam \rhos \;, \\
H^2&=& \frac{8\pi}{3 M_{Pl}^2}\left(\rho_{mm}+\rho_{r}+\rho_{dm}\right)\;,
\eea
where the subscript ``mm'' stands for mother matter, the component responsible for the early period of matter domination of the Universe, ``dm'' for daughter matter and $f_b$ is the fraction of mother matter decaying into daughter matter.

Within this kind of Universe, one can reconstruct the history of the Universe by taking suitable values for both $\gam$ and $f_b$. These two parameters are not independent of each other but can be related as follows 

\bea
f_b&\simeq& \frac{T_{\rm eq}}{T_{\rm RH}} \; ,\label{eq.fb}\\
T_{{\rm RH}}&\simeq& 0.55g_*^{-1/4}\sqrt{M_{Pl}\gam}\;,
\eea
where $T_{\rm RH}$ is the reheating temperature, {\it i.e.}, the temperature at which the mother particle releases all its energy and $T_{\rm eq}$ is the temperature when the energy density of radiation and matter are equal, with $T_{\rm eq} \simeq 1 \ {\rm eV}$ to get the right amount of dark matter today.

Therefore, if we require that the mother particle has completely decayed away prior to BBN, one may obtain a lower bound on $\gam\gtrsim2.0\times10^{-24} \ {\rm GeV}$ or likewise an upper bound on $f_b \lesssim 10^{-6}$. Such small values of the branching fraction might be dangerous for the formation of mini-halos. As it was demonstrated by Cen \cite{Cen:2000xv}, the density of mini-halos $\rho_{\rm mini-halo}$ decreases by a factor $(f_b)^4$ when a sizeable portion of the main component of such substructures decays into radiation
\beq
\rho_{\rm mini-halo}|_f= (f_b)^4 \rho_{\rm mini-halo}|_i\; ,
\eeq
where the subscript ``i'' referes to values before the decay and ``f'' after the decay is completed. Hence, for the mini-halos not to puff up by a large factor, the branching ratio $f_b$ to non-relativistic particles ({\it i.e.} to the daughter particles) cannot be very small. %, since the density of the mini-halo will decrease by a factor $(f_b)^4$. 
Consequently, one would need $f_b$ to be not far from unity in order for the substructure to survive. However, as in this scenario the daughter particle is the only dark matter component, then such large values for the branching fraction $f_b$ are forbidden by current observations, as they would lead to an overabundance of dark matter today. Therefore we need to add new channels of entropy production. 

\begin{figure}[t]
 \centering
 \includegraphics[scale=0.65]{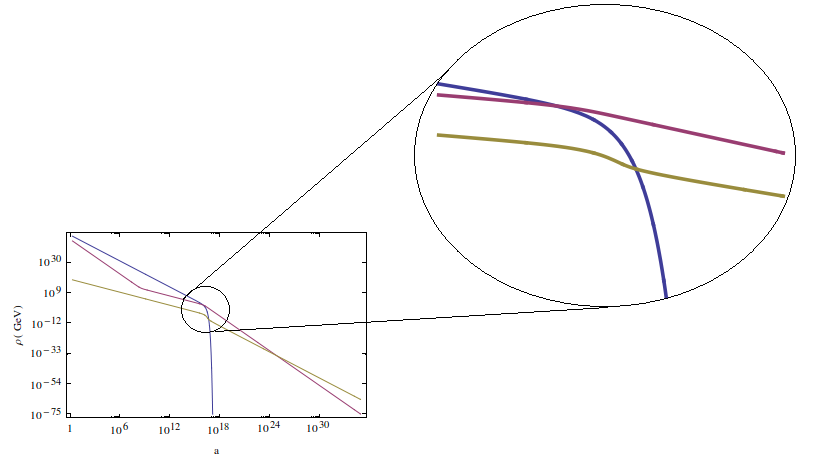}
 % thermal history.png: 794x1123 pixel, 96dpi, 21.00x29.70 cm, bb=0 0 595 842
 \caption{Evolution of the energy densities for the different components of our Universe. The transition between the early matter domination epoch and the standard radiation one is depicted amplified}
 \label{fig:thermalhistory}
\end{figure}

Owing to the mentioned argument, our setup will be the same as exposed earlier but including now a channel of annihilation for the daughter matter into radiation. The equations of motion for the energy densities can be then written as follows
\bea
\frac{\drm\rho_{\rm mm}}{\drm t}  +3H\rho_{{\rm mm}} &=& -\gam \rhos\;, \\
\frac{\drm\rho_r}{\drm t} +4H\rho_r &=& (1-f_b)\gam \rhos + \Upsilon^{anh}\;, \\
\frac{\drm\rho_{dm}}{\drm t} +3H\rho_{dm} &=& f_b\gam \rhos - \Upsilon^{anh}\;, \\
H^2&=& \frac{8\pi}{3 M_{Pl}^2}\left(\rho_{mm}+\rho_{r}+\rho_{dm}\right)\;, 
\eea
where
\beq
\Upsilon^{anh}(t)=\gamma \left(\rho_{dm}^2(t)-\rho_{\rm eq}^2(t)\right)\nonumber\\
\eeq
is the operator for the annihilation of daughter matter into radiation. $\rho_{eq}(t)$ is the equilibrium  density. Given the fact that our daughter matter density is not a thermal relic, we have set this density to 0.

The motivation for including such a term is clear. As it was showed before, mini-halos formation throughout a period of matter domination decaying into radiation are suppressed by the fourth power of $f_b$. Therefore, any substructure formed would be erased given the upper bound from eq.(\ref{eq.fb}) unless we include this new channel that alleviates this effect, letting more production of daughter matter and allowing us to have $f_b$ as large as needed. Particularly, one might take $f_b$ equal to 1, a situation in which the annihilation would be the only source of all the radiation in the Universe\footnote{This particular case however would require a reheating temperature of several hundred GeVs or higher. As we will see, as the size of the objects formed during this early matter domination era obviosuly depends on the length of this era, this  extremal case with $f_b\sim 1$ is clearly not favoured}. 

In addition to the conditions explained above, one must not only care about the fraction of matter produced during this period, but also about the velocity at which they are expelled after being formed since perturbations might be washed out by great velocities through free streaming. For such an analisis we need to define a scale $\lambda_{fs}$ \cite{Bertschinger:2006nq}
\beq
\lambda_{fs}(t)=\int_{t_{reh}}^t\frac{<v_{dm}>}{a} \ dt
\eeq
below which perturbations get erased away.

If one assumes that the velocity after reheating is diluted linearly by the expansion of the Universe and that the free streaming scales barely change after matter-radiation equality, one finds that \cite{Boyarsky:2008xj}
\bea
\lambda_{fs}(a)&=&\frac{<v_{RH}> a_{RH}}{H_0 \sqrt{\Omega_{rad}}}\int_{a_{reh}}^a\frac{1}{a'\sqrt{1+a_{eq}/a'}} \ da'\nonumber\\
&=&\frac{2 <v_{RH}> a_{RH}}{H_0 \sqrt{\Omega_{rad}}}\left(\arcsinh{\sqrt{\frac{a_{eq}}{a_{RH}}}}-\arcsinh{\sqrt{\frac{a_{eq}}{a}}}\right)\;,
\eea
where $<v_{RH}>\equiv<v_{dm}(a_{RH})>$ is the average velocity of the daughter particle at the reheating moment (within the instantaneous decay approximation), $H_0$ is the current Hubble constant and $\Omega_{rad}$ the observed current abundance of radiation. 

Regarding the velocity of the daughter particle and assuming that one mother produces a pair of daughters, it can be then easily demonstrated by kinematics that
\beq
v_{dm}^2=\left(1-\frac{4m_{dm}^2}{M^2}\right)\; ,
\eeq
where $v_{dm}$ is given in units of $c$, $M$ is the mass of the mother particle and $m_{dm}$ the mass of the daughter matter. Consequently, one can see that in order for the daughter particle to have low velocities when created, so that to avoid free streaming washout effects, it needs to be nearly half of the mass of the mother particle.

On the other hand, we still need the daughter particle to give rise to the right amount of observed dark matter. Once the mother matter decays away completely at $T_{\rm RH}$, the density of daughter matter can be written as $\rho_{\rm dm}\simeq f_b \rho_{\rm rad}$. At this point, the annihilation term dominates over the expansion term in the equation for $\rho_{\rm dm}$, making it decay abruptly until both terms balance. This effect takes place shortly after the reheating time, where $\rho_{\rm dm}\simeq \frac{H(a_{\rm RH})}{\gamma}$. From that point onwards, the remaining density dilutes in the standard way with the expansion of the Universe to provide the observed amount of matter. This allows us to constrain the size of the annihilation coupling to be
\beq\label{eq.sizanh}
\gamma\simeq 5\times10^{-1} \frac{1}{\left(M_{Pl} \ T_{\rm RH} \ T_{\rm eq}\right)}=5,5\times 10^{-8}\left(\frac{1 \ {\rm MeV}}{T_{\rm RH}}\right) \quad {\rm GeV}^{-3}\;.
\eeq

In summary, the conditions/ingredients any model leaving traces of an earlier epoch of matter domination should have, are the following

\begin{enumerate}
 \item A heavy particle, that we have called ``mother'',  dominates the energy density of the Universe up to its decay into radiation and matter. The latter one is labeled as ``daughter''.
\item The daughter particle would be the candidate for WIMP dark matter.
\item The heavy mother particle forms mini-halos during the first matter dominated era before it decays. In order for those structures not to  evaporate completely, the daughter particles must be borned non- 
relativistic, {\it i.e.}, their masses must be nearly degenerate with that of the mother ($m\approx \frac{M}{2}$).
\item For the mini-halos not to puff up by a large factor, the  
branching ratio $f_b$ to non-relativistic particles ({\it i.e.}, to the  
daughter particle) cannot be very small since the density of the mini-halo decreases by a factor $(f_b)^4$. 
\item On the other hand, in order to have a radiation dominated universe  
by the time of BBN and until the usual epoch of matter domination,  
$f_b$ cannot be extremely large if it is the only source of entropy production.
\item It is impossible to simultaneously satisfy the last two conditions unless one includes a annihilation term, whose size is constrained by equation \ref{eq.sizanh}.
\end{enumerate}

\section{Perturbations and structure formation}

From the previous section, we have seen that it is plausible to have an Universe dominated by a very heavy particle which finally decayed into radiation and common matter. Moreover, as it was also pointed out, density perturbations entering the horizon during that epoch, can grow significantly until the non-linear regime is reached and form substructures. These substructures are very sensitive to the production of entropy, so high abundances of radiation during this epoch may delete any substructure formed. Therefore, we added an annihilation term for the daughter matter, which will mainly act after the heavy particle decayed away  allowing to have less amount of radiation during this early matter domination epoch. 

\begin{figure}[t]
 \centering
 \includegraphics[]{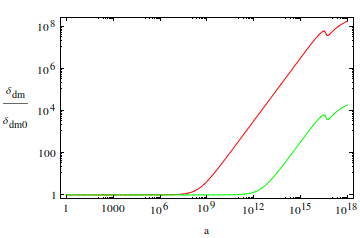}
\caption{Evolution in scale factor of the density contrast of the daughter particle, our would be dark matter candidate, for two different scales in units of the initial perturbation. Red line corresponds to $k=10^{4}k_{RH}$ and green line to $k=100k_{RH}$. Structures become non linear for $(\delta_{dm}/\delta_{dm0})\sim 10^5$ corresponding to $\delta_{dm0} \sim 10^{-5}$ as seen by CMB measurements. The arbitrary initial value for the scale factor has been taken equal to 1 when solving the equations of motion.}
 % ddm.png: 360x247 pixel, 72dpi, 12.70x8.71 cm, bb=0 0 360 247
\label{fig.ddm}
\end{figure}

The equations for the density perturbations read as follows
{\small
\begin{subequations} 
\begin{align}
a^2 E(a) \dels^\prime(a) +\thest(a)+3a^2E(a)\Phi^\prime(a) &= a \gamt \Phi(a)\;, \\
a^2 E(a) \thest^\prime(a)+aE(a)\thest+\tilk^2\Phi(a) &=0\;, \\
a^2 E(a) \delr^\prime(a)+\frac{4}{3}\thert(a)+4a^2 E(a) \Phi^\prime(a)&=  (1-f)\frac{\rhost^0(a)}{\rhort^0(a)}a\gamt\left[\dels(a)-\delr(a)-\Phi(a)\right] \ + \nonumber\\
 & +\frac{a}{H_1}\frac{\gamma}{\rho_r}\left[\left(\rho^2_{dm}-\rho_{eq}^2\right)(\delta_r + \Phi)-2\delta_{dm}\rho_{dm}^2\right] \label{dradanh}\;,\\
a^2 E(a) \thert^\prime(a)+\tilk^2\Phi(a) -\tilk^2\frac{\delr(a)}{4}&= (1-f)\frac{\rhost^0(a)}{\rhort^0(a)}a\gamt\left[\frac{3}{4}\thest(a)-\ther(a)\right] \ +\nonumber\\&  + \frac{a}{H_1} \frac{\gamma\left(\rho^2_{dm}-\rho_{eq}^2\right)}{\rho^0_r}\left[-\frac{3}{4}\theta_{dm}+\theta_r\right]\;,\\
a^2 E(a) \delm^\prime(a) +\themt(a)+3a^2E(a)\Phi^\prime(a) &=  f\frac{\rhost^0(a)}{\rhomt^0(a)}a\gamt\left[\dels(a)-\delm(a)-\Phi(a)\right] \ + \nonumber\\ & + \frac{a}{H_1} \left( - \frac{\gamma}{\rho_{dm}}\right)\left[\left(\rho^2_{dm}-\rho_{eq}^2\right)(\delta_{dm} + \Phi)-2\delta_{dm}\rho_{dm}^2\right]\;,  \displaybreak \\
a^2 E(a) \themt^\prime(a)+aE(a)\themt+\tilk^2\Phi(a) &=  f\frac{\rhost^0(a)}{\rhomt^0(a)}a\gamt\left[\thest(a)-\themt(a)\right]\label{thetadmanh} \;, \\
\tilk^2\Phi +3aE^2(a)\left[a^2\Phi^\prime(a)+a\Phi(a)\right] &= \frac{3}{2}a^2\left[\rhost^0(a)\dels(a)+\rhort^0(a)\delr(a)+\rhomt^0\delm(a)\right]\;,\nonumber\\
\end{align}
\end{subequations}}
where $E(a)\equiv\frac{H(a)}{H_0},\ \tilk\equiv\frac{k}{H_0}, \ \tilde{\theta}_{\{mm,dm,r\}}\equiv\frac{\theta_{\{mm,dm,r\}}}{H_0}$ and $\tilde{\rho}_{\{mm,dm,r\}}\equiv\frac{\rho_{\{mm,dm,r\}}}{\rho_0}$ with $H_0$ and $\rho_0$ being the initial Hubble rate and total energy density of the Universe respectively. The details about the derivation of these equations are given in appendix \ref{sec.pertbwc}. 

The equations given above reproduce the ones in \cite{Erickcek:2011us}, once the annihilation terms are set to zero. It can be seen that these terms source the equations for the density perturbations and velocities of the daughter and radiation component respectively (eqs. (\ref{dradanh}-\ref{thetadmanh})), playing a fundamental role when the mother component is on the verge of decaying and allowing the daughter particle to release part of its energy into radiation. This entropy production can be seen in figure \ref{fig.ddm}, where perturbations in the energy density of the daughter particle, {\it i.e.} what would be our dark matter candidate, are depicted and we can see that they tend to decrease several orders of magnitude when the mother particle decays away completely. Fortunately, as it was already mentioned, this decrease takes place near the reheating point, so it is expected that any density perturbations which have already entered in the non-linear regime will survive although their size can slightly decrease. In addition, one should notice that in both figures, what is plotted is the density contrast, {\it i.e.} the size of the perturbations in terms of its initial size. Such a quantity is determined by Inflation but anisotropy measurements in the CMB map set it to be  $\delta_{dm0}\approx 10^{-5}$, which in the standard case of adiabatic perturbations, takes this value for all the components of the Universe\footnote{We are assuming that the nearly scale invariance of primordial perturbations from Inflation still holds for such small scales}. Therefore, the non-linear regime would be reached in our figures when $\delta_{dm}$ times the size of the seeds is of order one. 

\begin{figure}[t]
 \centering
 \includegraphics[]{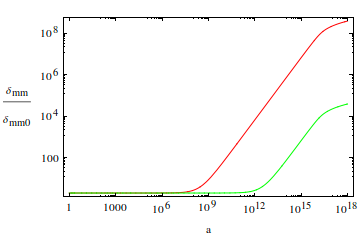}
\caption{Evolution in scale factor of the density contrast of the mother particle for two different scales in units of the initial perturbation. Red line corresponds to $k=10^{4}k_{RH}$ and green line to $k=100k_{RH}$. Structures become non linear for $(\delta_{mm}/\delta_{mm0}) \sim 10^5$ corresponding to $\delta_{mm0} \sim 10^{-5}$ as seen by CMB measurements. The arbitrary initial value for the scale factor has been taken equal to 1 when solving the equations of motion.} % ddm.png: 360x247 pixel, 72dpi, 12.70x8.71 cm, bb=0 0 360 247
\label{fig.dphi}
\end{figure}

Likewise, it is important to clarify that even the smallest scales which enter the horizon during this early matter domination epoch and do not have enough time to reach the non-linear regime (the example in figure \ref{fig.ddm} with $k=100k_{RH}$), {\bf  will still experience a remarkable growth, which will lead to the formation of structures and substructures much earlier than in the standard picture once the Universe becomes matter dominated again.} In particular, this feature may be important for a Dark Matter Halo to collapse shortly after matter-radiation equality forming an ultracompact minihalo, which are excellent indirect detection targets \cite{UCMH} and attractive for lensing prospects \cite{Ricotti:2009bs}.

In figure \ref{fig.dphi} the evolution of the density perturbations of the mother particle is shown. As it can be seen, such an evolution, behaving as matter, is very similar to that of the daughter except for the effect coming from the annihilation channel, which is absent in this component.

\begin{figure}[t]
 \centering
 \includegraphics[]{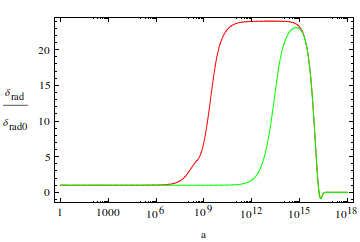}
\caption{Evolution in scale factor of the density constrast of radiation for two different scales. Red line corresponds to $k=10^{4}k_{RH}$ and green line to $k=100k_{RH}$. The initial contrast value corresponds to $\delta_{rad0}\simeq 10^{-5}$ as seen by CMB experiments. The arbitrary initial value for the scale factor has been taken equal to 1 when solving the equations of motion.}
 % ddm.png: 360x247 pixel, 72dpi, 12.70x8.71 cm, bb=0 0 360 247
\label{fig.drad}
\end{figure}

Finally, the evolution of the radiation perturbations are plotted in figure \ref{fig.drad}. As we can see, the amplitude is amplified during the early matter domination epoch until it decreases completely and begins to oscillate with a negligible value when the mother particle releases all the energy. Such behaviour is very similar to the one given in \cite{Erickcek:2011us}, a fact which exhibits that the annihilation channel has a minor effect on radiation perturbations.

Regarding the density perturbations and structure formation, a variable to study is $\sigma$, the variance of the density perturbations smoothed at a certain scale, normally used to analise, within the Press-Schechter formalism \cite{Press:1973iz}, the abundance and evolution of halos and sub-halos at relevant scales and with a certain size, which is given by

\beq\label{eq.sigma}
\sigma_{{\rm RH}}^2(R)=\int_0^\infty \frac{dk}{k} \left(\frac{k}{a_{\rm RH} H(a_{\rm RH})}\right)^4 W^2(kR)T^2(k)\delta_H^2(k)\;,
\eeq
where the subscript RH means that this quantity is evaluated at the reheating time when the mother particle releases all the energy.

Let us briefly explain the formula (\ref{eq.sigma}). As it was already mentioned, $\sigma^2$ is the density perturbations smoothed for a certain scale $R$. This role is played by the function $W(kR)$, which is responsible for filtering out those modes with $kR \geq 1$ and therefore allow us to study the relevant scales. In order to do this, we have used the following filter function
\beq
W(kR)=\exp(-\frac{1}{2}k^2(\alpha R)^2)\times W_{\rm top-hat}(kR)\;,
\eeq
where $W_{\rm top-hat}(kR)=\frac{3}{(kR)^3}\left[\sin(kR)-(kR)\cos(kR)\right]$ is the usual top-hat window function. For our purposes, however, we wish to focus upon the scales that enter the horizon during the early matter domination epoch and this is not achieved with the usual top-hat window function. Owing to this, we introduced an exponential function to suppress modes with $k< k_{RH}$.

\begin{figure}[t]
\centering
 \includegraphics[scale=0.8]{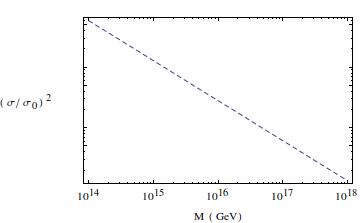}
 % sigma.png: 360x246 pixel, 72dpi, 12.70x8.68 cm, bb=0 0 360 246
 \caption{The variance of the daughter density perturbations $\sigma$ in terms of the mass $M$ contained in a sphere $R$. $\sigma_0$ is the normalization factor accounting for the several constants appearing in the equation \ref{eq.sigma}. The scale in both axes is logarithmic.}
 \label{fig:sigma}
\end{figure}

On the other hand, $T(k)$ is the well known transfer function which for the scales that we are taking into account, is scale invariant \cite{Erickcek:2011us} and $\delta_h(k)$ is the amplitude of the primordial density perturbations originated during inflation, which can be written as 
\begin{equation}
 \delta_h(k)=1.87\times10^{-5}\left(\frac{k}{k_{pivot}}\right)^{\frac{(n_s-1)}{2}}\; \;,
\end{equation}
where $k_{pivot}=0.002 \ {\rm Mpc}^{-1}$ and $n_s$ is the spectral index $n_s=0.9603\pm 0.0073$ \cite{Ade:2013uln}.

Finally, the factor $\left(\frac{k}{a_{\rm RH} H(a_{\rm RH})}\right)^4$ takes into account the scale factor growth of modes entering during matter domination.

In figure \ref{fig:sigma}, we show the normalised $\sigma^2$ (in arbitrary units) evaluated at the reheating epoch for different mass objects, related to their size by $\rho=\frac{M}{\frac{4\pi}{3}R^3}$, where $\rho$ is the total energy density at that moment. %and the values of $M$ correspond to objects whose components have physically reasonable mass.
 As it can be seen, this quantity is a mass decreasing function, meaning that the population of heavier objects is lower since they correspond to scales that entered later in the horizon and thus had less time to become non linear and begin to accrete matter.

The size of the dark matter objects formed is very model dependent, however to get a flavour of it, it is worth remembering that  the equivalent horizon mass scale at the QCD epoch ( $T\sim 100 \ {\rm MeV}$)  is around the mass of Jupiter. Moreover, one can easily work out the comoving Hubble size at the reheating time in terms of current parameters as 
\beq
k_{RH}^{-1}\sim 10^{-6} \sqrt{\frac{\Omega_r}{\Omega_m}} \left(  \frac{1 \ {\rm MeV}}{T_{RH}}\right)k_0^{-1}\;.
\eeq

Plugging the today known parameters, one can find that if the first matter domination era ends before BBN, a scale size which corresponds to roughly a parsec,  then the significant   power enhancement ({\it i.e}. formation of non-linear structure with perturbation amplitude of unity during the first matter domination era) would be on somewhat smaller scales than that, presumably corresponding to the milliparsec regime or even planets and stellar masses. 

\section{Conclusions}
In this work, we have shown that the formation of observational objects can be very sensitive to changes in the thermal history of the Universe. In particular, an early period of matter domination could amplify the primordial inflationary seeds leading to the formation of halos or mini-halos, objects which can be in principle observable and detectable \cite{UCMH, halos}.

Since at some point by BBN one needs to recover the usual picture of a radiation dominated Universe, one needs to care about the transition between both phases due to the production of entropy. Such production may erase or at least reduce any structure formed during the early period of matter. In particular, the density of primordial objects is suppressed by the fourth power of the branching fraction into radiation of the leading component during the matter epoch. As one needs to connect this scenario with the usual picture, {\it i.e.} radiation domination by BBN and right amount of dark matter abundance, one is forced to using values of the branching ratio which dilutes any primordial objects.

In order to solve this, we have introduced a new channel for the annihilation of the daughter matter into radiation. This allows us to have less amount of radiation during the period of structure formation and thus, larger values of the branching function. Furthermore, as it was showed in the profile of figure \ref{fig:thermalhistory}, this new channel only plays an important role when all the energy of the mother particle is totally released, connecting the end of the early matter domination era with the usual picture, and therefore any dilution can only take place when perturbations have already entered in the non-linear regime. We have also showed that this will only happen for modes that entered the horizon early enough to fall into the non-linear regime. In terms of the scale factor, it will happen for $\left(a_{RH}/a\right)\gtrsim 10^6$ or $\left(k/k_{RH}\right) \gtrsim 10^3$. Modes with $1000 k_{RH}> k> k_{RH}$, {\it i.e.} that don't reach the non-linear regime before the decay of the mother particle, enter the non-linear regime in the second  matter domination epoch, but may start to collapse into potential wells much earlier than within the standard thermal history picture due to the earlier growth.

We have also estimated that the new objects beginning to form during this first matter dominated epoch correspond to the milliparsec regime. Can such small scales have any observational relevance for the CMB?. In principle, it is hard to tell since one needs to evolve the perturbations after they entered in the non-linear regime all the way throughout the radiation dominated epoch. Certainly, there are many intriguing features and potentially interesting signatures for models with a (long enough) early period of matter domination able to leave potentially observable substructures. Of course a complete analysis needs to be performed by making use of non-linear methods such as N-Body simulations and falls beyond the scope of this manuscript. Hopefully, our work will trigger such an analysis and above all will let the reader judge himself the grade of apprehension that is appropiate when examining the phenomenology of these theories that take us away from the standard thermal history of the Universe. 

 To make this picture complete, one may argue that the today existing dark matter abundance does not come primarily from the decay of the daughter particle but from the freeze-out of non relativisitic matter from thermal equilibrium. This would require smaller branching fractions $f_b$, with the consequent creation of even more entropy which would erase more easily the substructures formed. Moreover, given the features of many proposed candidates for dark matter in the freeze-out scenario (specially neutralino), one would need larger reheating temperatures which would then lead to a less prolonged period of matter domination. As it is at this epoch when perturbations can grow until they enter in the non-linear regime, a dark matter relic density coming only from the decay of a heavy particle appears to be the most favourable scenario regarding an early structure formation in the Universe.

Finally, one may wonder how the observed baryon asymmetry is generated in a scenario like this. At first sight, it seems that it can only come from the decay of the mother particle, imposing more restrictions on its properties. A mechanism viable with having such a heavy particle could be a net baryon number production by means of a derivative coupling of the mother particle to the lepton/baryon current. Such an operator yields an effective chemical potential for baryons and anti-baryones when CPT is violated, allowing the velocity of the heavy particle to develop a non-zero vacuum expectation value \cite{derivBar}. Alternatively, one could also resort to the electroweak phase transition to produce the baryon asymmetry by changing the underlying  thermal history of the Universe to being matter dominated during the EWPT, which requires an efficient baryogenesis mechanism due to the entropy production \cite{Barenboim:2012nh}.

%%%%%%%%%%%%%%%%%%%%%%%%%%%%%%%%%%%%%%%%%%%%%%%%%%%%%%%%%%%%%%%%%%%%%%
\section*{Acknowledgments}
%%%%%%%%%%%%%%%%%%%%%%%%%%%%%%%%%%%%%%%%%%%%%%%%%%%%%%%%%%%%%%%%%%%%%%
It is a pleasure to thank Scott Dodelson and Will Kinney for useful comments. The authors 
acknowledge financial support from spanish MEC and FEDER (EC) under grant 
FPA2011-23596, and Generalitat Valenciana under the grant PROMETEO II/2013/017. GB acknowledges partial support from the European Union FP7 ITN INVISIBLES (Marie Curie Actions, PITN- GA-2011- 289442).

\appendix
\section{Pertubation equations without annhiliation}\label{sec.pertbwoc}
In this section, we derive the perturbation equations when an operator for the decaying of the mother particle is added. 

In general, the energy conservation equation can be written in a covariant way as follows
\beq
\nabla_\mu\left(^{(i)}T^\mu_{ \nu}\right) = Q_\nu\;.
\eeq
For the case of decaying matter we have the following
\bea
Q_\nu^{(\phi)} &=& ^{(\phi)} T_{\mu\nu} u_\phi^\mu\Gamma_\phi\;,\\
Q_\nu^{(r)} &=& -(1-f)Q_\nu^{(\phi)}\;,\\
Q_\nu^{(dm)} &=&-f Q_\nu^{(\phi)}\;,
\eea
where $f_b$ is the branching fraction, $\Gamma_\phi$ is the decay operator, $u_\phi^\mu=\left(1-\psi,\vec{V}\right)$ is the perturbed 4-velocity and $T_{\mu\nu}$ is the stress energy-momentum tensor, which in the perfect fluid case reads
\beq
T^{\mu\nu}=(\rho + P) u^\mu u^\nu + Pg^{\mu\nu}\;.
\eeq

We shall work in the Newtonian gauge of the perturbed FRW metric, which reads as 
\beq
\drm s^2 = -(1+2\Psi)\drm t^2 + a^2(t)\delta_{ij}(1+2\Phi)\drm x^i \drm x^j\;.
\eeq

With the above ingredients, one is able to derive the perturbation equations \cite{Erickcek:2011us}

%\begin{subequations}
{\small 
\begin{align}
a^2 E(a) \dels^\prime(a) +\thest(a)+3a^2E(a)\Phi^\prime(a) &= a \gamt \Phi(a), \\
a^2 E(a) \thest^\prime(a)+aE(a)\thest+\tilk^2\Phi(a) &=0, \\
a^2 E(a) \delr^\prime(a)+\frac{4}{3}\thert(a)+4a^2 E(a) \Phi^\prime(a)&= (1-f)\frac{\rhost^0(a)}{\rhort^0(a)}a\gamt\left[\dels(a)-\delr(a)-\Phi(a)\right], \nonumber\\
\\
a^2 E(a) \thert^\prime(a)+\tilk^2\Phi(a) -\tilk^2\frac{\delr(a)}{4}&= (1-f)\frac{\rhost^0(a)}{\rhort^0(a)}a\gamt\left[\frac{3}{4}\thest(a)-\ther(a)\right], \\
a^2 E(a) \delm^\prime(a) +\themt(a)+3a^2E(a)\Phi^\prime(a) &= f\frac{\rhost^0(a)}{\rhomt^0(a)}a\gamt\left[\dels(a)-\delm(a)-\Phi(a)\right], \\
a^2 E(a) \themt^\prime(a)+aE(a)\themt+\tilk^2\Phi(a) &=  f\frac{\rhost^0(a)}{\rhomt^0(a)}a\gamt\left[\thest(a)-\themt(a)\right]\;,\\
\tilk^2\Phi +3aE^2(a)\left[a^2\Phi^\prime(a)+a\Phi(a)\right] &= \frac{3}{2}a^2\left[\rhost^0(a)\dels(a)+\rhort^0(a)\delr(a)+\rhomt^0\delm(a)\right],\nonumber\\
\end{align}}
%\end{subequations} 
where $E(a)\equiv\frac{H(a)}{H_0},\ \tilk\equiv\frac{k}{H_0}, \ \tilde{\theta}_{\{mm,dm,r\}}\equiv\frac{\theta_{\{mm,dm,r\}}}{H_0}$ and $\tilde{\rho}_{\{mm,dm,r\}}\equiv\frac{\rho_{\{mm,dm,r\}}}{\rho_0}$ with $H_0$ and $\rho_0$ being the initial Hubble rate and total energy density of the Universe respectively.

\section{Pertubation equations with annhiliation}\label{sec.pertbwc}

We will now focus on the modification of the density perturbation equations when including an annihilation term. 

If we now add a source term accounting for the annhiliation of matter into radiation, these equations would be given as
\begin{align}
Q_\nu^{(\phi)} &= ^{(\phi)} T_{\mu\nu} u_\phi^\mu\Gamma_\phi\;,\\
Q_\nu^{(r)} &= -(1-f)Q_\nu^{(\phi)}+Q_\nu^{anh}\;,\\
Q_\nu^{(dm)} &=-f Q_\nu^{(\phi)}-Q_\nu^{anh}\;.
\end{align}

So our ansatz for the annihilation source could be the following
\beq
Q^{anh}_\nu=-\gamma\left(^{(dm)}T_{\nu}^{\ r} \ ^{(dm)}T_{r\mu}-\rho_{eq}^2 \ g_{\nu\mu}\right)u^\mu\;.
\eeq

\begin{figure}[t]
 \begin{center} 
 \includegraphics[scale=0.6]{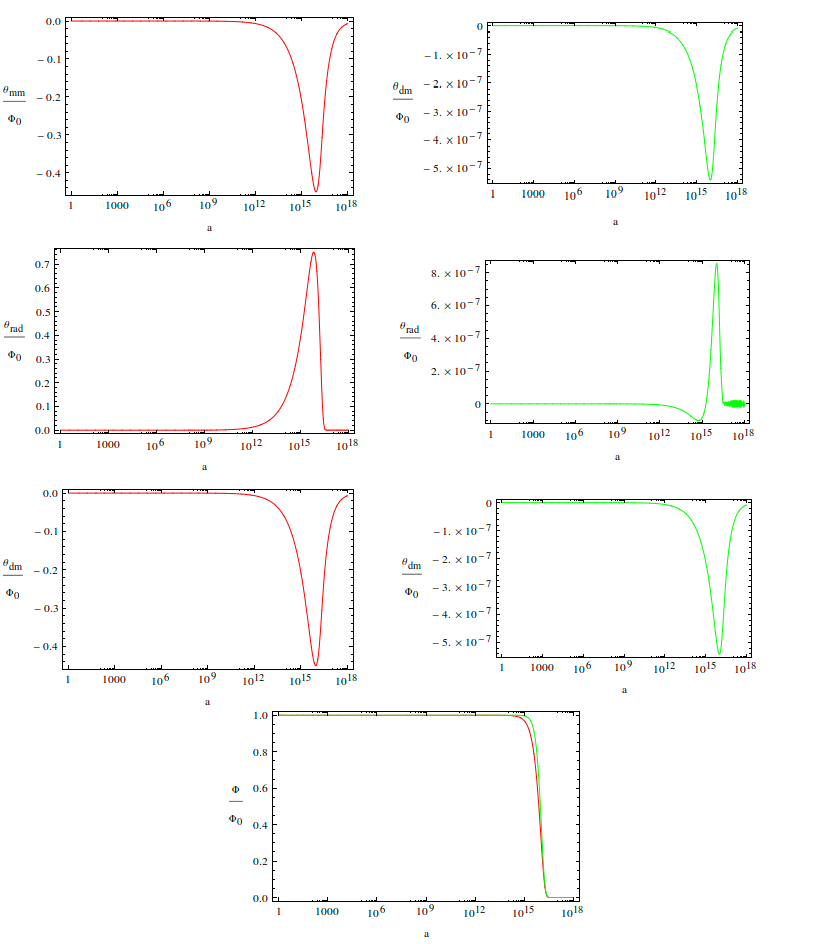}
\caption{Evolution in scale factor of the velocity components of the mother particle, radiation and daughter particle perturbations for two different scales. Red line corresponds to $k=10^{4}k_{RH}$ and green line to $k=100k_{RH}$. On the bottom it is pictured the evolution of the gravitational potential for the same pair of scales}
\end{center}
 % ddm.png: 360x247 pixel, 72dpi, 12.70x8.71 cm, bb=0 0 360 247
\label{fig.vel_pot}
\end{figure}

It can be seen that the zero component at zero order gives rise to the right operator

\begin{align}
Q^{anh}_0&=-\gamma\left(g_{0\lambda}\ ^{(dm)} T^{\lambda r} g_{rs}g_{\mu w} \ ^{(dm)}T^{sw}-\rho_{eq}^2 \ g_{0\mu}\right)u^\mu\nonumber\\
&=-\gamma\left(g_{0\lambda}\ \rho^{dm} u^{\lambda}u^{r} g_{rs}g_{\mu w} \rho^{dm} u^{s}u^{w} -\rho_{eq}^2 \ g_{0\mu}\right)u^\mu\nonumber\\
&=-\gamma\left( g_{0\lambda}\ \rho^{2,dm}(u_{s}\cdot u^{s}) u^{\lambda} (u_{\mu}\cdot u^\mu) -\rho_{eq}^2 \ g_{0\mu}u^\mu\right)\nonumber\\
&=-\gamma \ g_{00}\left(\rho^{2,dm}-\rho_{eq}^2\right) u^0\nonumber\\
&=+\gamma \left(\rho^{2,dm}-\rho_{eq}^2\right)\;,
\end{align}
where the relations $(u_s\cdot u^s)=-1$, $g_{00}= -1$ y $u^0=1$ have been used.

A first order in perturbations $Q_\nu^{inh}$ takes then the following form
\begin{align}
Q_0^{inh}&=\gamma\left[\left(\rho^{2,dm}-\rho_{eq}^2\right)(1+\Psi)+2\delta^{dm}\rho^{2,dm}\right]\label{eq.q0}\;,\\
Q_i^{inh}&=-a^2\gamma V_i\left(\rho^{2,dm}-\rho_{eq}^2\right)\label{eq.qi}\;.
\end{align}

Working out the energy conservation equation for each component with the perturbed metric in the Newtonian gauge, one can derive the perturbation equations but including now the annihilation terms. These equations read as follows
%\begin{subequations} 
{\small
\begin{align}
a^2 E(a) \dels^\prime(a) +\thest(a)+3a^2E(a)\Phi^\prime(a) &= a \gamt \Phi(a)\;, \label{delsa2}\\
a^2 E(a) \thest^\prime(a)+aE(a)\thest+\tilk^2\Phi(a) &=0\;, \label{thesa2}\\
a^2 E(a) \delr^\prime(a)+\frac{4}{3}\thert(a)+4a^2 E(a) \Phi^\prime(a)&= \dots +\frac{a}{H_1} \frac{1}{\rho_{r}^0}\left[Q_0^{anh,(0)} \delta_r -Q_0^{anh,(1)}\right] \;,\\
a^2 E(a) \thert^\prime(a)+\tilk^2\Phi(a) -\tilk^2\frac{\delr(a)}{4}&= \cdots + \frac{a}{H_1} \frac{1}{\rho^0_r}\left[\frac{\partial_i Q_i^{anh}}{a(1+w_{rad})} + Q_0^{anh,(0)}\theta_r\right]\;,\\
a^2 E(a) \delm^\prime(a) +\themt(a)+3a^2E(a)\Phi^\prime(a) &= \dots -\frac{a}{H_1} \frac{1}{\rho_{dm}^0}\left[Q_0^{anh,(0)} \delta_{dm} -Q_0^{anh,(1)}\right]\;,\\
a^2 E(a) \themt^\prime(a)+aE(a)\themt+\tilk^2\Phi(a) &=  \cdots +  0 \\
\tilk^2\Phi +3aE^2(a)\left[a^2\Phi^\prime(a)+a\Phi(a)\right] &= \frac{3}{2}a^2\left[\rhost^0(a)\dels(a)+\rhort^0(a)\delr(a)+\rhomt^0\delm(a)\right]\;, \nonumber\\
\end{align}}
%\end{subequations} 
\normalsize
where $(\dots)$ contains the terms without annihilation. If we show them explicitly, the equations of motions are written as follows
%\begin{subequations} 
{\small
\begin{align}
a^2 E(a) \dels^\prime(a) +\thest(a)+3a^2E(a)\Phi^\prime(a) &= a \gamt \Phi(a)\;,\\
a^2 E(a) \thest^\prime(a)+aE(a)\thest+\tilk^2\Phi(a) &=0\;, \\
a^2 E(a) \delr^\prime(a)+\frac{4}{3}\thert(a)+4a^2 E(a) \Phi^\prime(a)&=  (1-f)\frac{\rhost^0(a)}{\rhort^0(a)}a\gamt\left[\dels(a)-\delr(a)-\Phi(a)\right] \ + \nonumber\\
 & +\frac{a}{H_1}\frac{\gamma}{\rho_r}\left[\left(\rho^2_{dm}-\rho_{eq}^2\right)(\delta_r + \Phi)-2\delta_{dm}\rho_{dm}^2\right]\;, \\
a^2 E(a) \thert^\prime(a)+\tilk^2\Phi(a) -\tilk^2\frac{\delr(a)}{4}&= (1-f)\frac{\rhost^0(a)}{\rhort^0(a)}a\gamt\left[\frac{3}{4}\thest(a)-\ther(a)\right] \ +\nonumber\\&  + \frac{a}{H_1} \frac{\gamma\left(\rho^2_{dm}-\rho_{eq}^2\right)}{\rho^0_r}\left[-\frac{3}{4}\theta_{dm}+\theta_r\right]\;,\\
a^2 E(a) \delm^\prime(a) +\themt(a)+3a^2E(a)\Phi^\prime(a) &=  f\frac{\rhost^0(a)}{\rhomt^0(a)}a\gamt\left[\dels(a)-\delm(a)-\Phi(a)\right] \ + \nonumber\\ & + \frac{a}{H_1} \left( - \frac{\gamma}{\rho_{dm}}\right)\left[\left(\rho^2_{dm}-\rho_{eq}^2\right)(\delta_{dm} + \Phi)-2\delta_{dm}\rho_{dm}^2\right] \;,  
\\
a^2 E(a) \themt^\prime(a)+aE(a)\themt+\tilk^2\Phi(a) &=  f\frac{\rhost^0(a)}{\rhomt^0(a)}a\gamt\left[\thest(a)-\themt(a)\right] \;,\\
\tilk^2\Phi +3aE^2(a)\left[a^2\Phi^\prime(a)+a\Phi(a)\right] &= \frac{3}{2}a^2\left[\rhost^0(a)\dels(a)+\rhort^0(a)\delr(a)+\rhomt^0\delm(a)\right]\;,\nonumber\\
\end{align}}
%\end{subequations}

\end{document}